\input{aipcheck}

\documentclass[
    ,final            
  ]
  {aipproc}

\layoutstyle{6x9}

\def\lta{~\mbox{\raisebox{-.6ex}{$\stackrel{<}{\sim}$}}~}
\def\gta{~\mbox{\raisebox{-.6ex}{$\stackrel{>}{\sim}$}}~}
\def\lesssim{\lta}
\def\gtrsim{\gta}
\newcommand{\beq}{\begin{equation}}
\newcommand{\eeq}{\end{equation}}

\def\text{\rm}
\def\pla{{\rm P}}

\let\gam=w

\renewcommand{\epsilon}{\varepsilon}

\def\I{{\cal I}}

\newcommand{\gev}{\text{GeV}}
\newcommand{\ev}{\text{eV}}

\newcommand{\Mev}{\text{MeV}}
\newcommand{\s}{\text{s}}

\begin{document}

\title{R$^2$ Dark Matter\footnote{To appear in the Proceedings
of the sixth International Workshop on the Dark Side of the Universe
(DSU20010) Leon, Guanajuato, Mexico 1-6 June 2010.}}


\author{Jose A. R. Cembranos}{
  address={William I. Fine Theoretical Physics Institute,
University of Minnesota, Minneapolis, 55455, USA},
 altaddress={
School of Physics and Astronomy, University of Minnesota,
Minneapolis, 55455, USA} }

\begin{abstract}
There is a non-trivial four-derivative extension of the
gravitational spectrum that is free of ghosts and phenomenologically
viable. It is the so called $R^2$-gravity since it is defined by the
only addition of a term proportional to the square of the scalar
curvature. Just the presence of this term does not improve the
ultraviolet behaviour of Einstein gravity but introduces one
additional scalar degree of freedom that can account for the dark
matter of our Universe.
\end{abstract}

\maketitle


\section{Introduction}
\label{intro}

The non-unitarity and non-renormalizability of the gravitational
interaction described by the Einstein-Hilbert action (EHA) demands
its modification at high energies. It has been pointed out that this
correction cannot be accomplished without the introduction of new
states \cite{Cembranos:2008gj}; these states typically interact with
SM fields through Planck scale suppressed couplings and potentially
work as dark matter (DM).

In spite of many and continuous efforts, the ultraviolet (UV)
completion of the gravitational interaction is still an open
question. In these conditions, it is difficult to make general
statements about its phenomenology although different types of new
scalar fields are commonly predicted \cite{stgen,sft1}. We can adopt
a conservative and minimal approach in order to capture the
fundamental physics of this fact \cite{Cembranos:2008gj}. The
simplest correction to the EHA at high energies is provided by the
inclusion of four-derivative terms in the metric that preserve
general covariance. The most general four-derivative action
supports, in addition to the usual massless spin-two graviton, a
massive spin-two and a massive scalar mode, with a total of eight
degrees of freedom (in the {\it physical} or {\it transverse} gauge
\cite{Stelle:1976gc, Stelle:1977ry}). Indeed, four-derivative
gravity is renormalizable, although the massive spin-two gravitons
are ghost-like particles that generate new unitarity violations,
breaking of causality, and inadmissible instabilities
\cite{Simon:1991bm}.

\section{R$^2$ gravity}
\label{R2}

However, we can work with $R^2$-gravity, that is defined by the only
addition of a term proportional to the square of the scalar
curvature to the EHA. It illustrates the idea in a consistent and
minimal way since it only introduces one additional scalar degree of
freedom, whose mass $m_0$ is given by the corresponding new constant
in the action:
\begin{eqnarray}
\label{model} S_G&=&\int\sqrt{g}\left\{-\Lambda^4
-\frac{M_{\pla}^2}{2}R \,+\frac{M_{\pla}^2}{12\,m_0^2}R^2
\,+\,...\,\right\}
\end{eqnarray}
where $M_{\pla}\equiv (8\pi G_N)^{-1/2}\simeq 2.4 \times 10^{18}$
GeV, $\Lambda\simeq 2.3\times 10^{-3}$ eV, and the dots refer to
higher energy corrections that must be present in the model to
complete the UV limit. In \cite{Cembranos:2008gj}, it has been shown
that just the Action (\ref{model}) can explain the late time
cosmology since the first term can account for the dark energy (DE)
content, while the third term is able to explain the DM one.

$R^2$-gravity modifies Einstein's Equations (EEs) as
\cite{Starobinsky:1980te,Gottlober:1990um} (following notation from
\cite{Cembranos:2005fi}):
\begin{eqnarray}\label{equation1}
\left[1-\frac{1}{3\,m_0^2}\,R\right]R_{\mu\nu}-
\frac{1}{2}\left[R-\frac{1}{6\,m_0^2}\,R^2\right]g_{\mu\nu}
-\,\,\I_{\alpha\beta\mu\nu}
\nabla^\alpha\nabla^\beta\left[\frac{1}{3\,m_0^2}\,R\right]
=\frac{T_{\mu\nu}}{M_{\pla}^2}\,,
\end{eqnarray}
where $\I_{\alpha\beta\mu\nu}\equiv
\left(g_{\alpha\beta}g_{\mu\nu}-g_{\alpha\mu}g_{\beta\nu}\right)$.
The new terms do not modify the standard EEs at low energies except
for the mentioned introduction of a new mode. It is straight forward
to check that the metric $g_{\mu\nu}=[1+c_1 \sin(m_0 t)]
\eta_{\mu\nu}$ is solution of the linearized Eq. (\ref{equation1}),
i.e. for $c_1\ll 1$, without any kind of energy source. It has been
argue that the energy stored in such oscillations behaves exactly as
cold DM and can explain the missing matter problem of the Universe
\cite{Cembranos:2008gj}.

\section{Scalar graviton couplings}
\label{couplings}

The phenomenology of the new scalar can be computed inside the {\it
Jordan} or the {\it Einstein frame} by expanding the metric
perturbatively:
\begin{eqnarray}
\label{metricperturbation}
g_{\mu\nu}=\hat{g}_{\mu\nu}+\frac2{M_{\pla}}h_{\mu\nu}
-\sqrt{\frac23}\frac1{M_{\pla}}\,\phi\,\hat{g}_{\mu\nu}\,,
\end{eqnarray}
where $\hat{g}_{\mu\nu}$ is its classical background solution,
$h_{\mu\nu}$ takes into account the standard two degrees of freedom
associated with the spin-two (traceless) graviton, and $\phi$
corresponds to the new mode, which owns a standard kinetic term.

The couplings of this scalar graviton with the SM fields have been
computed in \cite{Cembranos:2008gj} by supposing that gravity is
minimally coupled to matter (in the Jordan frame). In such a case,
there is a linear coupling to matter through the trace of the
standard energy-momentum tensor:
\begin{eqnarray}
\label{coupling} {\cal
L}_{\phi-T_{\mu\nu}}&=&\frac{1}{M_{\pla}\sqrt{6}}\,\phi\,T^{\mu}_{\mu}\,.
\end{eqnarray}
Therefore, the couplings with the massive SM particles -Higgs boson
($\Phi$), (Dirac) fermions ($\psi$), and electroweak gauge bosons-
are:
\begin{eqnarray}
\label{mass}
{\cal L}^{\text{tree-level}}_{\phi-SM} &=&
\frac{1}{M_{\pla}\sqrt{6}}\,\phi\, \left\{2\,m_\Phi^2
\Phi^2-\nabla_\mu \Phi\nabla^\mu \Phi \right.
\\
&+& \sum_\psi m_\psi\, {\bar \psi} \psi \left.
 - 2\, m_W^2\, W^+_\mu {W^-}^\mu -\,m_Z^2\, Z_\mu Z^\mu
 \right\}\,.
\nonumber
\end{eqnarray}
In addition, this scalar graviton couples to photons and gluons due
to the {\it conformal anomaly} \cite{Cembranos:2008gj}: 
\begin{eqnarray}
\label{massless} {\cal L}^{\text{one-loop}}_{\phi-SM} &=&
\frac{1}{M_{\pla}\sqrt{6}}\,\phi\, \left\{ {\alpha_{EM} c_{EM} \over
8 \pi}\, F_{\mu\nu} F^{\mu\nu} \right.
+\left. {\alpha_s c_G  \over 8\pi}\, G^a_{\mu\nu} G_a^{\mu\nu}
\right\}\,.
\end{eqnarray}
The particular value of the couplings ($c_{EM}$ and $c_G$) depends
on the energy and possible heavy particles, charged with respect to
these gauge interactions, that may extend the SM at high energies.

\section{Scalar graviton abundance}
\label{abundance}

The thermal abundance that this field can achieve depends on the UV
completion of the theory. However, there is no reason to expect that
the initial value of the scalar field ($\phi_1$) should coincide
with the minimum of its potential ($\phi=0$) if $H(T)\gg m_0$. It
implies that the scalar graviton may have associated big abundances
through the so called {\it misalignment mechanism}. Below the
temperature $T_1$ for which $3H(T_1)\simeq m_0$, $\phi$ behaves as a
standard scalar. It oscillates around the minimum. These
oscillations correspond to a zero-momentum condensate, whose initial
number density: $n_\phi \sim m_0\phi_1^2/2$ (where
$\phi_1=\sqrt{\left\langle \phi(T_1)^2\right\rangle}$ ), will evolve
as the typical one associated to standard non-relativistic matter.
The abundance o this particle has been computed in
\cite{Cembranos:2008gj}:
\begin{eqnarray}
\Omega_{\phi}h^2&\simeq& 0.86\, \left[ \frac{m_0}{1\, \ev}
\right]^{\frac12} \left[ \frac{\phi_1}{10^{12}\, \gev} \right]^{2}
\left[ \frac{100\, g_{e\,1}^3}{(\gamma_{s1} g_{s1})^4}
\right]^{\frac14},\;\;\;\;\;\; \label{abundance}
\end{eqnarray}
where $g_{e\,1}$ ($g_{s1}$) are the effective energy (entropy)
number of relativistic degrees of freedom at $T_1$, $h\simeq 0.70$
is the Hubble parameter, and $\gamma_{s1}$ is the factor that this
entropy has increased in a comoving volume since the onset of scalar
oscillations. We see that initial conditions of order of $\phi_1\sim
10^{12}\;\gev$ can lead to the non-baryonic DM (NBDM) abundance
depending on the rest of parameters and the early physics of the
Universe (see Fig. \ref{gravDM}).

\section{Scalar graviton signatures}
\label{signatures}

On the other hand, Eqs. (\ref{coupling},\ref{mass}) imply that the
new scalar graviton mediates an attractive Yukawa force between two
non-relativistic particles of masses $M_a$ and $M_b$:
\begin{eqnarray}
\label{Yukawa} V_{ab}&=&-\frac{1}{24 \pi M_{\pla}^2}\frac{M_a
M_b}{r} e^{- m_0\,r}\,.
\end{eqnarray}
\begin{figure}[bt]
\resizebox{8.0cm}{!} {\includegraphics{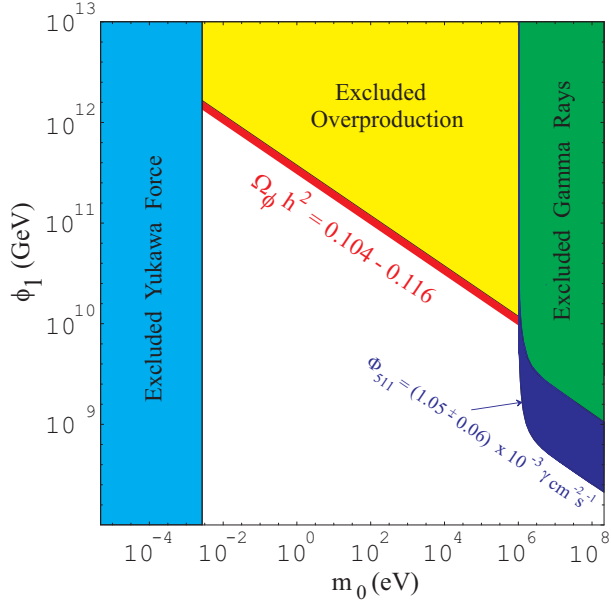}} \caption{Different
regions of the parameter space of R$^2$ gravity: $m_0$ is the mass
of the scalar graviton and $\phi_1$ is its misalignment when $3H\sim
m_0$ (we assume $g_{e\,1}=g_{s1}\simeq 106.75$, and
$\gamma_{s1}\simeq 1$). The left side is excluded by modifications
of Newton's law. The right one is excluded by cosmic ray
observations. In the limit of this region, $R^2$-gravity can account
for the positron production in order to explain the 511 keV line
coming from the GC confirmed by INTEGRAL \cite{Knodlseder:2005yq}
(up to $m_0 \sim 10\; \Mev$). The upper area is ruled out by DM
overproduction. The diagonal line corresponds to the NBDM abundance
fitted with WMAP data (Figure taken from \cite{Cembranos:2008gj}). }
\label{gravDM}
\end{figure}
Since it has not been observed, torsion-balance measurements are
able to constraint the scalar mass \cite{Cembranos:2008gj}:
\begin{eqnarray}
\label{yukawa}
m_0\geq 2.7 \times 10^{-3} \ev \;\;\;\;\;\;\;\text{at}\;\;95\%\;\;\text{c.l.}
\end{eqnarray}
This is the most important lower bound on this mass, and it is
independent of its abundance. Depending on its abundance, $m_0$ is
constrained from above. The decay in $e^+e^-$ is the most
constraining if $\phi$ constitutes the total NBDM. From
(\ref{mass}), it is possible to calculate the $\phi$ decay rate into
a generic pair fermion anti-fermion \cite{Cembranos:2008gj}.
Restrictions are set by the observations of the SPI spectrometer on
the INTEGRAL (International Gamma-ray Astrophysics Laboratory)
satellite, which has measured a 511 keV line emission of $1.05 \pm
0.06 \times 10^{-3}$ photons cm$^{-2}$ s$^{-1}$ from the Galactic
center (GC)~\cite{Knodlseder:2005yq}, confirming previous
measurements. This 511 keV line flux is fully consistent with an
$e^+ e^-$ annihilation spectrum although the source of positrons is
unknown.

If $m_0 \geq 1.2\; \Mev$, the scalar mode cannot constitute the
total local DM since we should observe a bigger excess of the 511
line coming from the GC. On the other hand, decaying DM (DDM) has
been already proposed in different works \cite{Picciotto:2004rp,
Cembranos:2008bw} as a possible source of the inferred positrons if
its mass is lighter than $M_{\text{DDM}}\lesssim 10\;\Mev$
\cite{Beacom:2005qv} and its decay rate in $e^+e^-$ verifies
\cite{Cembranos:2008gj}:
\begin{eqnarray}
\label{511} \frac{\Omega_{\text{DDM}}h^2\;
\Gamma_{\text{DDM}}}{M_{\text{DDM}}} \simeq \left[(0.2 - 4)\times
10^{27}\; \s\; \Mev\right]^{-1}
\,.
\end{eqnarray}
The most important uncertainty for this interval comes from the dark
halo profile, although a cuspy density is definitely needed (with a
inner slope $\gamma\gtrsim 1.5$ \cite{Cembranos:2008bw}). If $m_0$
is tunned to $2\,m_e$ with an accuracy of $5$-$10\,\%$, the line
could be explained by $R^2$-gravity. The same gravitational DM can
explain the 511 line with a less tuned mass (up to $m_0 \sim 10\;
\Mev$) if $\phi_1 \sim 10^9\; \gev$, i.e. with a lower abundance
(See Fig. \ref{gravDM}).

If $m_0 < 2\, m_e$, the only decay channel that may be observable is
in two photons. If $m_0\lesssim 1\; \Mev$, it is difficult to detect
these gravitational decays in  the isotropic diffuse photon
background (iDPB)~\cite{Cembranos:2007fj,Cembranos:2008bw}. However,
a more promising analysis is associated with the search of photon
lines at $E_\gamma= m_0/2$ from localized sources. The iDPB is
continuum since it suffers the cosmological redshift, but the
mono-energetic photons originated by local sources may give a clear
signal of $R^2$-gravity in future experiments if the scalar graviton
is inside the heavier allowed region of the model
\cite{Cembranos:2007fj}.

\section{Conclusions}
\label{conclusions}

Although there are other possibilities \cite{other}, DM is usually
assumed to be in the form of stable Weakly-interacting massive
particles (WIMPs) that naturally freeze-out with the right thermal
abundance. One of the most interesting features of WIMPs, is that
they emerge in well-motivated particle physics scenarios as in
R-parity conserving supersymmetry (SUSY) models~\cite{SUSY1,SUSY2},
universal extra dimensions (UED)~\cite{UED1, UED2}, or
brane-worlds~\cite{BW1,BW2,BW3}. In this analysis, we have studied
the possibility that the DM origin resides in UV modifications of
gravity. We have focused on $R^2$-gravity, but the low energy
phenomenology of the studied scalar mode is present in the same
well-motivated frameworks such as string theory, supersymmetry or
extra dimensional models (in form of dilatons, graviscalars of
radions). Another interesting property of WIMPs, it is that they can
be tested with high energy experiments as the new generation of
colliders \cite{Coll}. This possibility seems remote for the type of
gravitational DM discussed in this work. However, indirect
observations as modifications of Newton's law or cosmic rays can
provide the first signatures of this type of DM.

\begin{theacknowledgments}
DOE grant FG02-94ER40823, FPA 2005-02327 project
(DGICYT, Spain), CAM/UCM 910309 project, and MICINN
Consolider-Ingenio MULTIDARK CSD2009-00064.
\end{theacknowledgments}

\bibliographystyle{aipproc}   

\end{document}